\documentclass[12pt]{article}
\newcommand{\be}{\begin{equation}}
\newcommand{\ee}{\end{equation}}
\begin{document}
\baselineskip=24 pt
\begin{center}

{\large {\bf Extended Planck Scale}} 

\end{center}

\vskip1.5truecm 

\begin{center}
  F. I. Cooperstock 
 \\{\small \it Department of Physics and Astronomy, University 
of Victoria} \\
{\small \it P.O. Box 3055, Victoria, B.C. V8W 3P6 (Canada)}\\
V. Faraoni
\\

{\small \it Physics Department, University of Northern British
Columbia\\
3333 University Way, Prince George, B.C. V2N 4Z9 (Canada)}\\

{\small \it e-mail addresses: cooperstock@phys.uvic.ca, vfaraoni@unbc.ca}
\end{center} 
\begin{abstract} 
Traditional derivations of the Planck mass
ignore the role of charge and spin in general relativity. From the
Kerr-Newman null surface and horizon radii, quantized charge and spin dependence 
are introduced in an extended Planck scale of mass. Spectra emerge with 
selection rules dependent upon the choice of Kerr-Newman radius to link with the 
Compton wavelength. The appearance of the fine structure
constant suggests the possibility of a variation in time of the extended
Planck mass, which may be much larger than the variation in the
traditional one. There is a suggestion of a connection with the $\alpha$ value 
governing high-energy radiation in Z-boson production and decay.
\end{abstract} 

\begin{center}
\mbox{ }\\
\end{center}

Different arguments have led researchers to a measure of the scale at
which gravity must necessarily mesh with quantum theory, the
\textit{Planck scale}. The most common approach is to form a combination
of the gravitational constant $G$, the reduced Planck constant $\hbar$,
and the speed
of light $c$ that has the dimension of mass and label it the
\textit{Planck mass} $m_{p}$
\begin{equation} \label{eqp1}
m_{p}= \sqrt{ \frac{\hbar c}{G} } \simeq 2.2 \cdot 10^{-5} \, \mbox{g} 
\end{equation}
or, equivalently, one can consider the Planck length $l_{p}= \sqrt{ G
\hbar/c^3} \simeq 1.6 \cdot 10^{-33}$~cm, the Planck time $t_{p}=l_p/c\simeq
5.4 \cdot 10^{-44}$~s, or the Planck energy $E_{p}=m_{p} c^2 \simeq 1.3 \cdot
10^{19}$~GeV.

However, this approach does not
distinguish between Newtonian gravity and general relativity, the preferred 
\textit{relativistic} theory of gravity. A more
illuminating argument reflecting both the quantum scale and the role of
general relativity derives from equating the Compton wavelength of a
particle of mass $m$, namely $\lambda_C =\hbar/m c $, with its
gravitational radius $r_S = 2Gm/c^2 $, the
radius of its event horizon as found from the Schwarzschild metric. This gives 
the same result apart from a factor $1/\sqrt{2}$. In following this procedure, 
what is made transparent is that the ``Planck particle''  so
derived is without spin or 
charge. However, spin and charge are the fundamental quantized aspects of 
matter. To exclude them is to ignore the important couplings that spin and 
electromagnetism have to gravitation. Therefore to be general, we consider what 
effect their consideration has on what we now designate as the \textit{extended 
Planck} (henceforth referred to as \textit{``plex"}) scale. In place of the 
spinless neutral Schwarzschild particle, we consider a particle endowed with 
spin and charge, again within the context of general relativity. The metric for 
a body of mass $m$, charge $q$ and angular momentum per unit mass $a$ is
the Kerr-Newman metric (with $c = G =1$)  
\cite{KerrNewman} 
\begin{equation} \label{eqp2} 
ds^2= \frac{sin^2{\theta}}{\rho^2} \left[
\left( r^2 +a^2 \right) d\phi - a \, dt \right]^2-\frac{D}{\rho^2} \left[ 
dt - a \sin^2{ \theta } \, d\phi \right]^2 +\frac{\rho^2}{D} \, dr^2 +
{\rho^2} d\theta^2  \;,
\end{equation} 
where 
\begin{equation} \label{eqp2a} 
D \equiv r^2 - 2mr +a^2 + q^2 \;, \;\;\;\;\;\;\;  \rho^2 \equiv r^2 + a^2
\cos^2\theta  \;.
\end{equation} 
The new gravitational radius is (we now restore the $c$
and $G$) \cite{KerrNewman} 
\begin{equation} \label{eqp3} 
r_{+} = \frac{G}{c^2} \left( m + \sqrt{m^2 - \frac{q^2}{G}  -
\frac{c^2}{G^2} \, a^2} \,\, \right) \; .
\end{equation} 

However, with spin and charge added, there is scope to focus on the other 
significant radius, the radius of the``null surface" $r_{-}$ with the
negative sign in front of the square root
\begin{equation} \label{eqp3a} 
r_{-} = \frac{G}{c^2} \left( m - \sqrt{m^2 - \frac{q^2}{G}  -
\frac{c^2}{G^2} \, a^2} \,\, \right) \; .
\end{equation} 
 
Since we are dealing with the quantum domain, we quantize
the charge in units of the charge $ e $ of the electron and
the angular momentum  in units of the fundamental quantum of angular
momentum  $\hbar$, with respective quantum numbers $N$
and $s$: 
\begin{equation} \label{eqp4} 
q=N \, e \;, \;\;\;\;\;\;\;  a=s \, \frac{\hbar}{m} \;. 
\end{equation} 
(Note that the $m$ appears again through the spin.) When one sets the
Kerr-Newman event horizon (eq.~(\ref{eqp3})) and null surface
(eq.~(\ref{eqp3a}))
radii of  the particles equal to their
Compton wavelengths, and substitutes the quantized charge and spin from
eq.~(\ref{eqp4}), one has
\begin{equation} \label{eqp4a}
\frac{\hbar}{mc} = \frac{G}{c^2} \left( m \pm \sqrt{m^2
-\frac{N^2 e^2}{G} - \frac{c^2 \hbar^2 \, s^2}{G^2m^2}} \,\, \right) \;.
\end{equation}
At this point in the nascent state of development of the subject, it is unclear 
whether it is $r_{+}$ or $r_{-}$ that should be the length scale to connect with 
the Compton wavelength in the quantum domain or indeed, if both values have a 
role to play. Accordingly, in what follows, both possibilities will be 
investigated.
Solving for $m$, one finds that the mass which we now refer to as the
\textit{extended Planck mass} $m_{plex}$ is
\begin{equation}  \label{eqp5}
m_{plex} = m_{pl} \, \sqrt{ \frac{2(1+s^2)}{2 -\alpha N^2 }} \; ,
\end{equation}
for both cases, where $\alpha \equiv  e^2/\hbar c \simeq 1/137 $ is the fine 
structure
constant
and henceforth we
use the subscript notation ``pl" to designate the standard Planck mass  
with the $1/\sqrt{2}$ factor included, $m_{pl} \equiv \sqrt{ \hbar c / (
2G)} = m_p /\sqrt{2} $.
By eq.~(\ref{eqp5}), the presence of either spin or charge leads to
an increase in the value of $m_{plex}$ as compared to the traditional
$m_{pl}$. Moreover, the presence of the fine structure constant in
eq.~(\ref{eqp5}) 
provides an additional source of interest, given the current focus upon its 
apparent slow variation in time \cite{Webbetal}-\cite{Murphyetal}. 

Following recent claims \cite{Webbetal}-\cite{Murphyetal} that the value
of the fine structure constant underwent changes during the last half of
the history of the universe, we focus on the possibility that
$\alpha$ could have had a considerably different value in the still more
distant  past. Although rather
unorthodox in the low-energy regime, this idea appears quite naturally in
the context of renormalization, in which the coupling ``constants'' are
actually running couplings. In the standard model, the early
universe expands and cools precipitously in its very first instants
when it emerges from the big bang, and the energy scale drops substantially, 
allowing for significant variations in the values of the running couplings.

It has been claimed that if the fine structure ``constant'' changes at all,
a change in $c$ rather than $e$ is responsible as a change in $e$ would
violate the laws of black hole thermodynamics \cite{Daviesetal}.  A
time-varying 
$\alpha$ can be accomodated in the
context of varying speed of light cosmologies, of which
many proposals have appeared recently \cite{VSL0}-\cite{VSL6} (see however the 
criticism in Ref.~\cite{Carlip}).
While the reported variation
of $\alpha $ over the last $10^{10}$  years is minute (of the order
of $10^{-5}$ \cite{Webbetal}-\cite{Murphyetal}) and the variation of
fundamental constants is restricted by
primordial nucleosynthesis, it is quite conceivable that more radical
changes could have occurred earlier in the history of the universe.
Although the 
current evidence points to a small increase in $\alpha$ as we go forward in time 
over the time scale thus far surveyed, the essential point is that there is 
variation and this variation could have been one of decrease from a larger value 
at a still earlier time. 
To fix our ideas, suppose that $N=5$ and $s$ is of order unity. Then, if
at sometime in the past, $\alpha$ assumed a value close to $ 8 \cdot 10^{-2}$
(approximately one order of magnitude larger
than its present value), the value of the extended Planck mass $m_{plex}$
would have
been many orders of magnitude larger than its present-day value,
regardless of
the value of the quantum number $s$ (larger values of $N$ lead to large
effects for smaller variations of $\alpha $). By contrast, if this change in the 
value of
$ \alpha $ was due to the time variation of $c$, the change in
the traditional Planck mass $m_{pl}$ instead would be relatively insignificant.
Since there has been some debate as to whether it is a variation in $e$ or
in $c $ 
that has been responsible for the observed change in $\alpha$, we point
out that 
the effects in the two cases upon the value of $m_{plex}$ are different. If it 
is $e$ that varies, this appears only in $\alpha$ in $m_{plex}$ whereas if
it is $c$ 
that is responsible, this change affects another part of the $m_{plex}$ 
expression as well.

Extremal values are generally useful to gain insight and hence 
it is perhaps worth noting that the critical upper-limit $N$ value in 
eq.~(\ref{eqp5}) 
is $N=16$ for the present $\alpha$ value of 1/137.036. With this $N$ 
value, the 
extended Planck scale becomes infinite for an $\alpha$ value of 1/128. 
Interestingly, the $\alpha$ value governing high-energy radiation in Z-boson 
production and decay has been measured to be 1/127.934, suggesting that there 
really may be some connection between fundamental constants and integers 
(recalling the history of theorizing about the number 137).

It is to be noted that the scope for the extension of the Planck scale is 
severely limited if one were to be restricted by the choice of the event
horizon 
radius eq.~(\ref{eqp3}) as opposed to the null surface radius
eq.~(\ref{eqp3a}). From  eq.~(\ref{eqp4a}) with the positive sign in front
of the square
root, one
finds the  inequality
\begin{equation} \label{eqp5a}
\frac{\hbar}{ mc} - \frac{Gm}{c^2} \geq 0
\end{equation} 
and hence, with eq.~(\ref{eqp5})
\begin{equation} \label{eqp5b}
m_{pl} \leq m_{plex} \leq \sqrt{2} \, m_{pl}
\end{equation}

These conditions in conjunction with eq.~(\ref{eqp5}) place the following 
restrictions on the allowed spin and charge quanta:

\begin{equation} \label{eqp7}
s^2 + N^2\alpha \leq 1 \;, \;\;\; \; \; \; \; N^2 \alpha < 2,
\end{equation}
Thus, the allowed values of $s$ and $N$ for $\alpha = 1/137$ are

a) for $s=0$, $ N \leq 11$

b) for $s=1/2$, $N \leq 10$ 

c) for $s=1$, $N=0$.
Note that spin two is not allowed in this case and this might evoke
some 
surprise as the graviton is seen as a spin two boson. However the extended
Planck 
mass, as the traditional Planck mass, is very large whereas the graviton
mass  is zero to a very high level of accuracy ($m_{graviton} <
10^{-59}$~g). They are  very different concepts.

Given the new extended approach, it is natural to introduce an extended {\em 
Planck 
charge} and a {\em Planck spin}. These quantities could be defined by assuming 
that the ``Planck
particle'' considered is an extremal black hole, i.e. one defined by
\be
m^2=\frac{q^2}{G} + \frac{c^2}{G^2}\; a^2 
\ee
(corresponding to the equality in (\ref{eqp7})) that is maximally
charged ($s=0$, $q=q_{max}$) or maximally rotating
($q=0$, $s=s_{max}$).  These requirements yield the extended Planck quantities
\be
q_{plex}=\frac{e}{\sqrt{\alpha}}\simeq 11.7 \, e \;, \;\;\;\;\;\;\;\;\;\;\;
s_{plex}=1 
\ee
(corresponding to the Planck angular momentum $L_{plex} = \hbar$ and now 
allowing for non-integral $N$).
While $q_{plex} $ is large but not extraordinarily so, $L_{plex}$ is rather 
ordinary on the scale of particles familiar at an energy much lower than the 
Planck
scale. This is the reason why the inclusion of charge and spin does not
appreciably change the value of the extended Planck mass $m_{plex}$ with respect 
to
$m_{pl}$ of the spinless, neutral case, if one assumes that $\alpha$ does
not vary. (See, however, below where the null surface radius is used to relate 
to the Compton wavelength.)

According to the third law of black hole thermodynamics, an extremal black
hole corresponds to zero absolute temperature, and is an unattainable
state. If the third law survives in the Planck regime, the
values of $N$ and $s$ are even further restricted, and the first of
(\ref{eqp7}) should read as a strict inequality.

If one considers instead the null surface of radius $r_{-}$ defined by
eqs.~(\ref{eqp3a}) and (\ref{eqp5}) 
the inequalities
\begin{equation} \label{eqp8}
s^2 + N^2\alpha \geq 1 \;, \;\;\;\;\;\;\;\;\; N^2\alpha <2
\end{equation}
follow.

In this case, the allowed values of $s$ and $N$ for $\alpha = 1/137$ are,

a) for $s=0$, $12 \leq N \leq 16$

b) for $s=1/2$, $11 \leq N \leq 16$

c) for $s=1$, $0 \leq N \leq 16$

d) for $s=2$, $0 \leq N \leq 16$

In this case, spin two is readily allowed.

Particle masses get renormalized and hence behave like running couplings. 
Perhaps this is the case as well for the extended Planck mass although this is 
speculation in the absence of a renormalizable theory of quantum gravity. 
Perhaps what we have in the substance of the extended Planck mass is a semi-
classical analogue of renormalization.

{\small {\bf Acknowledgments:} This work was supported in part by a grant
from the Natural Sciences and Engineering Research Council of Canada.}

\clearpage
{\small 
\end{document}